# Measurement Methodology for Determining the Optimal Frequency Domain Configuration to Accurately Record WiFi Exposure Levels

Marta Fernandez, *Student Member, IEEE,* David Guerra, *Member, IEEE*, Unai Gil, *Member, IEEE*, Ivan Peña, *Member IEEE*, Amaia Arrinda, *Senior Member, IEEE*

*Abstract*— **Radiofrequency fields are usually measured in order to be compared with electromagnetic exposure limits defined by international standardization organizations with the aim of preserving the human health. However, in the case of WiFi technology, accurate measurement of the radiation coming from user terminals and access points is a great challenge due to the nature of these emissions, which are non-continuous signals transmitted in the form of pulses of short duration. Most of the methodologies defined up to now for determining WiFi exposure levels use or take as reference exposimeters, broadband probes and spectrum analyzers without taking into account that WiFi signals are not continuously transmitted. This leads to an overestimation of the radiation level that cannot be considered negligible when data of the actual exposure are needed. To avoid this, other procedures apply empirical weighting factors that account for the actual duration of burst transmissions. However, this implies the implementation of additional measurements for calculating the weighting factors, and thus, increases the complexity of the work. According to this, it was still necessary to define the frequency domain measurement setup that is optimal for obtaining realistic WiFi signal values, without requiring the performance of additional recordings. Thus, the definition of an appropriate methodology to achieve this goal was established as the main objective of this study. The set of tasks carried out to identify such configuration, as well as the limitations obtained for other measurement settings are deeply explained in this paper.**

## I. Introduction

Human exposure to wireless signals is nowadays a matter on the spotlight of our society due to the current massive use of this type of communications, and therefore, a growing number of solutions developed in order to monitor electromagnetic fields (EMF), such as the ones presented in [1], [2] have been defined in the last years. In this regard, there is special concern about WiFi networks, since they are increasingly being deployed in both public and private areas for a wide range of applications, as the support of independent life of elderly people [3], or the monitoring of ambient conditions [4]. According to the review carried out by Foster et. al [5], different studies state the existence of potential risks and effects caused by the exposure to WiFi radiation. Consequently, EMF levels generated by wireless communication systems operating in different environments should be measured in order to check the compliance with the human exposure limits established by different international regulation bodies, among which the International Commission on Non-Ionizing Radiation Protection (ICNIRP) [6] and the Institute of Electrical and Electronics Engineers (IEEE) [7] can be emphasized.

Although there are guidelines and standards that provide general information and techniques for measuring EMFs, they do not provide specific information for the case of WiFi signals [6], [8]. One of the main drawbacks in this regard is that the accurate assessment of this type of radiation poses a challenge because of the quasi-stochastic nature of these emissions resulting from their transmission in the form of bursts. Inexpensive radiofrequency detectors can lead to misleading results [5], [9]. Moreover, the configuration of specialized equipment, such as spectrum analyzers, has significant influence on the obtained values [10]. In fact, the behavior in the frequency domain of different parameters of a spectrum analyzer when measuring WiFi signals was analyzed in [11], considering a power meter equipped with a broadband probe as a reference system. The use of these instruments is suitable for having a rough approximation of exposure that may be useful as long as the levels are well below the exposure limits. Hence, the accuracy of the measurements can be improved if a more appropriate reference system is taken. As stated in [12] and [13], broadband probes do not provide enough accuracy for measuring the radiation caused by Orthogonal Frequency Division Multiplexing (OFDM) communication systems, and thus, it is necessary to employ other instruments when the objective of the measurements is to obtain realistic values to be used, for example, in medical studies carried out for the characterization of the influence of specific signal levels on the human body, or in the design of network planning methodologies according to criteria based on the actual exposure conditions.

Different methods for measuring human exposure due to WiFi signals with a spectrum analyzer have been defined.

This paragraph of the first footnote will contain the date on which you submitted your paper for review. It will also contain support information, including sponsor and financial support acknowledgment.

The authors are with the Communications Engineering Department, University of the Basque Country, 48013 Bilbao, Spain (e-mail: martafernandez010@gmail.com; david.guerra@ehu.eus; unai.gil@ehu.eus; ivan.pena@ehu.eus; amaia.arrinda@ehu.eus).

Nevertheless, there is still no standardized methodology for this purpose. The most common technique considered up to now is to record maximum power values in the frequency band of interest, so as to analyze the worst-case scenario. However, this also implies an overestimation of the radiation, which could derive in overly restrictive deployment policies because of social concern. To avoid this, some authors introduced weighting techniques that account for the time variability of these emissions. In [10] an empirical factor called *Duty Cycle* was defined as the ratio of the pulse duration to the total duration of the Wireless Local Area Network (WLAN) signal, in order to consider the time variability of the signal in a specific situation. Also, the spectrum analyzer settings for measuring maximum signal levels that should be subsequently weighted by that factor were proposed in that study. Another approach that takes into account both the amplitude and time variability of the received signals was described in [14], [15]. In that case, the weighting factor was determined as the ratio of the time-averaged power level to the maximum power level of the signal.

Those techniques were the solution adopted in different measurement campaigns carried out to assess human exposure to WiFi emissions. For example, the electric field coming from access points and portable devices when doing different activities was analyzed from recordings taken by means of exposimeters [16], [17] or other frequency selective radiation meters [18]. Max-hold WiFi measurements given by a spectrum analyzer were studied for different environments in [19]. In [20] different values of the Duty Cycle were considered to correct the empirical measurements. Nevertheless, the methodologies defined in all the previous studies were not optimal. In the case of portable exposure meters, uncertainties due to different factors such as the body influence have been reported in [21]. Moreover, the maximum values of WiFi exposure do not reveal a realistic situation and, although these maximum levels can be corrected by using weighting factors, two types of recordings are required in that case: one in the frequency domain to determine those maximum levels, and another in the time domain to fix the proper weighting value for the characteristics of the environment and users' activity under test.

Bearing in mind the problems derived from the previous methodologies, a rigorous procedure to identify the optimal configuration for determining realistic WiFi exposure values by acquiring samples only and exclusively in the frequency domain – an exposure value requiring less than 1 second to be recorded – was established as the main objective of this work. Such procedure has been based on both time and frequency domain measurements. The first type of measurements was necessary to obtain a set of reference samples of the radiation caused by a perfectly known WiFi signal. Once obtained those reference samples, they were compared with the levels registered for the same type of signal considering different values of the spectrum analyzer parameters in order to analyze their influence on the measurements. Finally, the optimal configuration was identified from recordings taken for different cases of WiFi reception. Thus, according to the objectives, work and results of this study the contents of the paper have been distributed as follows. The criteria and results utilized for the selection of the measurement instrument are described in Section II. The procedure defined to determine the optimal setup for measuring realistic WiFi exposure values only and exclusively in the frequency domain is explained in Section III. The experimental data derived from the application of that procedure and the configuration adopted as solution are presented in Section IV. To conclude, a methodology for taking WiFi exposure samples based on the configuration proposed as solution in this paper and the conclusions and suggestions for carrying out further work in this regard have been included respectively in Sections V and VI.

## II. Selection of The Measurement Instruments

The measurement instruments that are commonly used to acquire values of the exposure to EMFs are broadband probes, exposimeters and spectrum analyzers. As stated in the introduction, the first ones do not provide enough accuracy for measuring the radiation caused by OFDM communication systems, and thus, they were discarded for the acquisition of WiFi radiation samples in the frequency domain. Among exposimeters and spectrum analyzers, it is logical to assume that the second ones are the best option for recording values of the radiation caused by different types of signals, in different environments and under very different conditions, since most of the models include several parameters and options that confer them great measurement versatility. Even so, a set of tests carried out to compare the accuracy of a spectrum analyzer connected to a tri-axial antenna with the one provided by a portable exposimeter were initially performed as part of the tasks carried out in this study to identify the optimum solution for obtaining realistic WiFi exposure values.

The specific models utilized to do this were the EMI ESPI3 spectrum analyzer of Rohde & Schwarz [22] and the EME Spy 200 exposimeter of Satimo [23]. Both of them were selected because they fulfill the specifications defined for the professional equipment to be used for exposure assessment. Nevertheless, the conclusions derived from the results included not only in this section, but also in the following ones are applicable to a great variety of models such as Agilent E4402B or Agilent E443A from Keysight technologies [24], MS2840A from Anritsu [25] or FSC from Rhode and Schwarz [26] in the case of spectrum analyzers and ExpoM-RF from fields at work [27] or ESM-140 from Maschek [28] in the case of exposimeters, since according to their data sheets all of them present similar measurement characteristics.

Fig. 1 shows the differences for a set of WiFi field strength samples recorded with both instruments at the same time, separated each other a distance of 40 cm in order to ensure the corresponding far field conditions. Ten measurements of 6-minute duration were taken at two different positions (in total, twenty measurements with each equipment) and the electric field levels were averaged over these 6 minutes, as recommended by the ICNIRP [6].

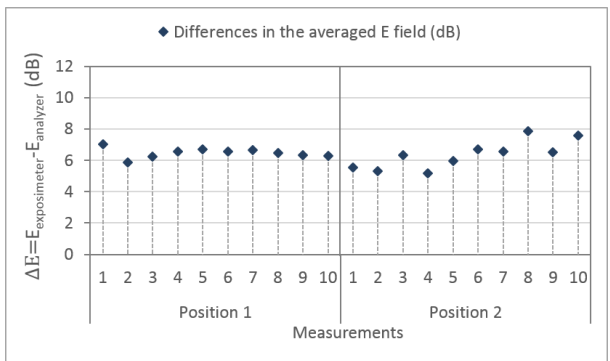

Fig. 1. Differences of the electric field levels measured with the exposimeter and the spectrum analyzer.

As deduced from the curves depicted in that figure, the levels measured with the exposimeter were always higher than the ones recorded with the analyzer. This is in part because the minimum detection threshold of the exposimeter is 5 mV/m, and consequently, that was the value stored by this instrument when lower levels were received. On the contrary, the spectrum analyzer not only captured levels below that threshold, but also recorded samples four times faster than the EME Spy, concluding that it is a better option to obtain empirical values of the radiation caused by amplitude and time varying WiFi signal bursts.

Nevertheless, it has to be considered that the accuracy of the spectrum analyzer is directly related with the configuration used during the measurements. Authors of this paper proved in [29] that in the case of WiFi exposure measurements of 6 minute duration, the values recorded when using the max-hold trace can be up to 15.1 dB higher than the ones obtained by means of a clear/ write trace together with the RMS detector. The reason is that only the maximum field strength levels received during the measurement time are stored when using a max-hold configuration which, in fact, is not suitable when the aim is to determine realistic exposure values accounting for the transmission/reception of short duration pulses, as in the case of WiFi networks. Therefore, the comparison between the exposimeter and the analyzer previously described was done using the clear/ write trace and the RMS detector of the spectrum analyzer. However, other parameters, such as the resolution bandwidth (RBW), video bandwidth (VBW) and sweep time (SWT) have also influence on the accuracy of the results. Having concluded that an analyzer configured to use the clear/ write trace together with the RMS detector and connected to a tri-axial antenna is a more accurate and versatile measurement solution than an exposimeter, the following step was to identify the optimum configuration for acquiring realistic WiFi exposure values by performing measurements only and exclusively in the frequency domain. To do this, a comparison of a set of reference levels measured in the time domain for a perfectly known WiFi signal with the ones obtained for the same type of signal with different frequency domain configurations was required. Details in this regard, as well the corresponding results and conclusions are given in the following sections.

## III. Measurement Methodology

The time and frequency domain measurements required to determine the optimal configuration of the spectrum analyzer for measuring accurate WiFi exposure levels in the frequency domain were performed in a laboratory of the University of the Basque Country (Spain), where a Cisco Aironet 1702 Access Point [30] provides access to the Eduroam WiFi network of the university. Tests were carried out at night from 00:00 to 05:00 in order to guarantee that no one was within the coverage area of that access point and that all the computers located in that area were turned off, ensuring this way that the only traffic received was the one produced for the testing. Furthermore, during the post-processing phase, all the recordings were analyzed to check that the received signals matched exactly the WiFi activity desired for the trials. Specifically, for those time and frequency domain measurements carried out to analyze the levels recorded when the access point was working in idle mode, it was confirmed that, as described in the IEEE 802.11 standard [31], only beacons were received periodically, while in the rest of cases, bi-univocal correspondence was observed between the signal traces and the data bursts particularly generated for tests. The access point operated on Channel 1 of the frequency band allocated to services that implement the IEEE 802.11g standard. The center frequency and bandwidth of this channel are 2.412 GHz and 20 MHz respectively. Also, it comprises 64 subcarriers equally spaced 312.5 kHz and modulated by applying the OFDM technique, so that the signal is concretely transmitted on subcarriers -26 to -1 and 1 to 26, being the 0 subcarrier the one located at the center frequency previously mentioned, as described in [31].

According to the IEEE Standard C95.3 [32], the measurement of potentially hazardous exposure fields coming from a well-known single-source may be performed with a tunable field-strength meter connected to a directive antenna, which in fact makes full sense, since in that case the exposure levels come from the specific location where that source is placed. Thus, bearing in mind that the objective of the study here described was to define a measurement methodology for determining the optimal settings to record actual WiFi radiation values and that it was necessary to use different WiFi signals coming from a specific access point for achieving that purpose, a Yagi antenna suitable for carrying out measurements at the 2.4 GHz WiFi band was selected as appropriate solution to be connected to the spectrum analyzer. Such antenna was installed on a mast, aiming at the access point from a distance of 2 m in order to receive as much power as possible from the radiation source under study. Besides, a computer using a wired Ethernet connection to the analyzer was employed to establish the analyzer configuration and save the recorded data. A laptop was used to generate data traffic from the access point.

During the trials, several data files of three different sizes were downloaded from a server that was part of the same local area network as the access point, so Internet traffic constraints were negligible. Also, two working modes of the access point were considered: the *idle mode*, where only beacon packets are transmitted, and the *traffic mode*, in which apart from the beacons data traffic is generated. Details of the specific tests performed for each mode, distinguishing the three measurement phases of the methodology defined as objective of this paper are given in the following subsections.

*Phase 1: Acquisition of reference WiFi exposure values*

The purpose of the first trials carried out with the measurement system previously described was to obtain a set of samples of the radiation caused by a perfectly known WiFi signal to be used as reference values for determining the optimal frequency domain configuration of the spectrum analyzer.

To do this, the power levels of the signal generated by the access point when working in idle mode were recorded in the time domain using the configuration indicated in the second column of Table I.

TABLE I
SPECTRUM ANALYZER CONFIGURATIONS

| Parameter | Time Domain | Frequency domain |
|---|---|---|
| fc (MHz) | 2412 ± 0.3125N<br>N= 0, 1, 2, … 32 | 2412 |
| Span (MHz) | Zero Span | 20 |
| RBW (MHz) | 0.3 | 0.3 - 1 |
| VBW (MHz) | 1 | 1 - 3 |
| SWT (s) | 1 | $2.5 \times 10^{-3} - 40 \times 10^{-3}$ |
| SWP | 8001 points | 501 points |
| Detector | RMS | RMS |
| Trace Mode | Clear/Write | Clear/Write |

Measurements in the frequency domain were discarded during this phase of the methodology since, as depicted in Fig. 2, the idle mode signal consisted of a sequence of beacons of 0.5-ms duration transmitted every 50 ms, and according to the spectrum analysis basics reported in [33], it is impossible to configure an analyzer for sweeping a WiFi channel in that short period of 0.5 ms, without losing the tradeoff between the values of the SWT, Span and RBW required to perform accurate recordings.

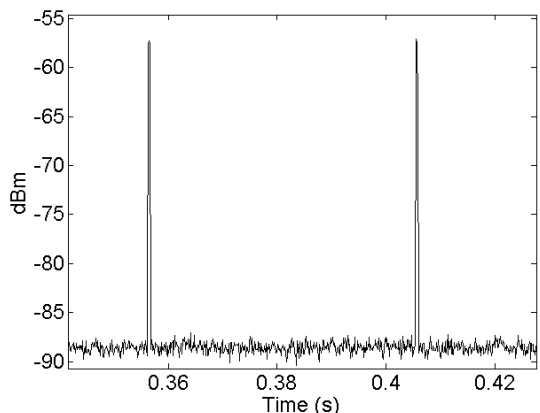

Fig. 2. Signal transmitted at 2.415 GHz by the access point when working in idle mode.

For this reason, the channel power $P_{channel}$ was calculated from the power levels recorded at the different frequencies $P_i$ in the time domain using equation (1), which makes the integration of the $P_i$ values as the spectrum analyzer does when it calculates the channel power from the displayed values at the different frequencies [33]. These measurements were taken separately at 65 different frequencies within the WiFi channel, recording data during intervals of one hour at each frequency.

$$P_{channel} = \frac{CHBW}{RBW} \cdot \frac{1}{N} \cdot \sum_{i=1}^{N} P_i \qquad (1)$$

where both $P_{channel}$ and $P_i$ are the before mentioned power values in linear units, $CHBW$ is the channel bandwidth (20 MHz), $RBW$ is the resolution bandwidth (0.3 MHz) and $N$ is the number of frequencies within the WiFi channel at which samples were recorded ($N$=65).

Fig. 3 shows the values registered at each one of the frequencies measured within the channel whenever the beacons were received, that is, every 50 ms. These values correspond with the maximum power levels recorded during the 1-hour duration measurements, so that the channel power obtained after converting to dBm the result calculated using equation (1) was -43.07 dBm in that case.

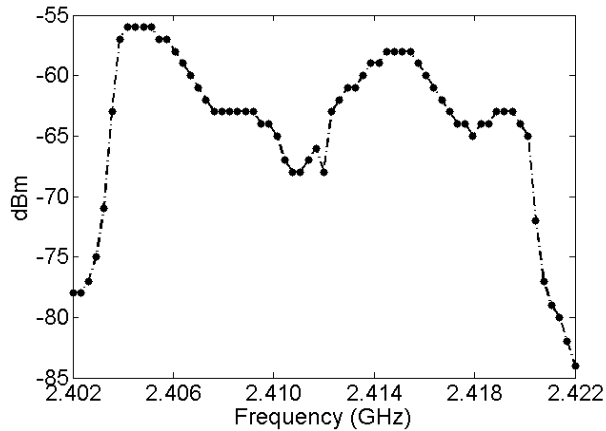

Fig. 3. Power levels registered at each one of the frequencies measured within the WiFi channel whenever the beacons were received.

*Phase 2: Study of the influence of the measurement parameters*

Although the most accurate way of characterizing the WiFi exposure is to use the power levels obtained by means of the above-mentioned procedure, it is not practical due to the amount of recordings that are required. Moreover, that method is not applicable to perform actual traffic measurements when changing the transmission conditions, since the different time domain measurements cannot be taken at the same time. However, it is the most accurate methodology for obtaining reference power values to assess the accuracy of other techniques. Bearing this in mind, the following objective was to determine the influence of the spectrum analyzer parameters on the measurements, in order to identify the optimal frequency domain configuration for registering the power levels that best fit the reference channel power values calculated by applying the time domain measurement method previously described.

To do this, samples of the signal transmitted by the access point when working in idle mode were taken in the frequency domain at the center frequency (fc) of 2.412 GHz, by using the

RMS detector and the Clear/Write trace, while varying the resolution bandwidth (RBW), the video bandwidth (VBW) and the sweep time (SWT) of the spectrum analyzer. It was observed that 501 points were enough to display that signal. That is, a greater amount of points did not improve the results, and thus, this was the value selected for the Sweep Points (SWP) parameter of the analyzer.

According to this, the RBW had to be larger than 40 kHz, as this was the separation between the displayed points (i.e., the Span value divided by the SWP value) [34]. This fixed also the minimum VBW since, as described in the CENELEC EN 50492 Recommendation, the value of this parameter should be at least 3 times higher than the RBW [35].

Taking into account the previous thresholds, the measurements carried out during this second phase were finally performed varying the RBW between 0.3 MHz and 1 MHz, the VBW between 1 MHz and 3 MHz and the SWT between 2.5 ms and 40 ms, as indicated in the third column of Table I. Results of these tests are included in Section IV-A of this document and led to conclude that the SWT is the most influential parameter when measuring WiFi emissions, due to the variability of this type of signals; an effect that was also confirmed in [10]. If the sweep time is too large, two or more beacons can be considered as a unique larger beacon, which indeed would imply an overestimation of the signal.

Nevertheless, although the effect caused by the SWT parameter was clearly observed during the measurements, the error associated with each one of the frequency domain configurations was quantified by applying the following equation, in order to make an objective comparison between them:

$$e\ (\%) = \frac{|P_{channel} - P_{freq}|}{|P_{channel}|} \cdot 100 \qquad (2)$$

where $e$ is the error, $P_{channel}$ is the power value obtained for the idle mode signal from the reference samples recorded in the time domain and $P_{freq}$ is the one obtained for the same signal with a specific frequency domain configuration of the spectrum analyzer, both of them expressed in linear units.

*Phase 3: Identification of the spectrum analyzer optimal configuration*

Apart from the recordings of the signal transmitted by the access point when working in idle mode, measurements of WiFi signals derived from different data traffic situations were required to identify the optimal configuration of the spectrum analyzer for acquiring actual WiFi exposure values only and exclusively in the frequency domain. To do this, files of three different sizes were downloaded from a server operating in the same local area network as the access point under test. Data traffic was generated by using a laptop located at a distance of 4 m from the receiver. Frequency domain measurements were taken in different intervals of 6-minute duration, as proposed by the ICNIRP Guidelines [6]. Specifically, the recording technique was the one following described: each measurement started when the access point was in idle mode, one minute later data traffic was generated and when reaching the 6 minutes, the spectrum analyzer stopped recording samples. Thus, the synchronization of the two software tools designed, on one hand, to download the files from the server, and on the other hand, to configure the analyzer and save the results, was essential for the success of these tests.

Again in this case, different values of the spectrum analyzer parameters shown in Table I, were considered to perform this third set of measurements with the objective of studying the relationship between the periods of time in which data traffic was generated and the power levels obtained with each configuration. An example of one of the traces recorded when the access point was working in traffic mode can be seen in Fig. 4. In this specific case the resolution bandwidth was set to 0.3 MHz.

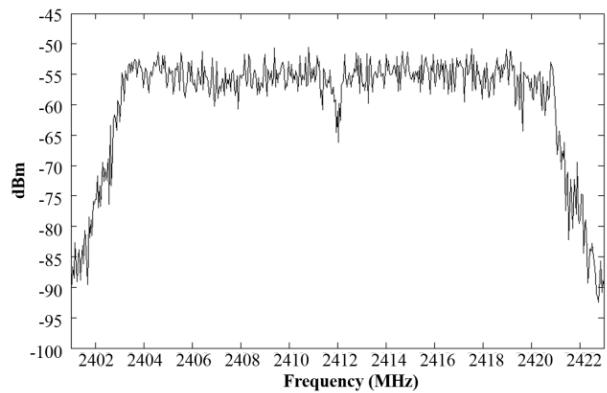

Fig. 4. Trace of the WiFi signal recorded in the frequency domain when downloading a data file.

In view of such WiFi activity, it was necessary to determine the percentages of signal reception during the 6-minute measurement intervals. Data downloading took between 2 s and 6 s for File 1, between 50 s and 60 s for File 2 and between 228 s and 240 s for File 3. Under such conditions, the access point was transmitting almost continuously, and therefore, taking into account that the duration of each measurement was six minutes, and considering that beacons were transmitted during the 1% of the time (that is, 0.5 ms each 50 ms) when data traffic was not generated, the time percentage of WiFi reception was 2-3% for the first type of file, 15-18% for the second type and 64-68% for the third one. These values led to conclude that the three file types were suitable to generate WiFi exposure situations different from each other, as well as different from the situation where the access point worked in idle mode.

## IV. Results

This section includes an analysis of the power levels and error values obtained from the sets of measurements described in the previous section that finally led to determine the optimal setup of the spectrum analyzer for acquiring realistic WiFi exposure values in the frequency domain. To do this, the results were classified according to the types of WiFi signals recorded during the tests.

*A. Results obtained for the idle mode signal*

The cumulative distribution functions (CDFs) calculated from the power values measured in idle mode in the frequency domain were compared with the CDF of the channel power values obtained from the samples recorded for that type of signal in the time domain during the phase 1 of the measurement methodology explained in the previous section. This last one is the so-called "Reference" curve depicted in black color in Fig. 5a and Fig. 5b. As mentioned in Section III, the maximum power value of that curve (that is, the highest reference power value) was -43.07 dBm. Besides, as seen in those figures, this curve drops to -65.89 dBm for the 99th percentile, and to -69.05 dBm for the 97th percentile, reaching finally a minimum level equal to -71.40 dBm.

CDFs corresponding to the measurements carried out in the frequency domain by using a resolution bandwidth of 0.3 MHz, a video bandwidth of 1 MHz and SWT values of 2.5 ms, 10 ms, 25 ms and 40 ms have been also included in Fig. 5a, while a set of curves corresponding to frequency domain measurements performed with the same SWT values, but RBW and VBW values of 1 MHz and 3 MHz respectively, can be observed in Fig. 5b. In all these cases, data were stored during time periods of 1 hour. Moreover, different tests of 1-hour and 6-minute duration were performed, concluding that there was no difference in this regard. Thus, as stated in [6], an interval of 6 minutes can be considered long enough to determine the exposure level, if the environment conditions remain constant.

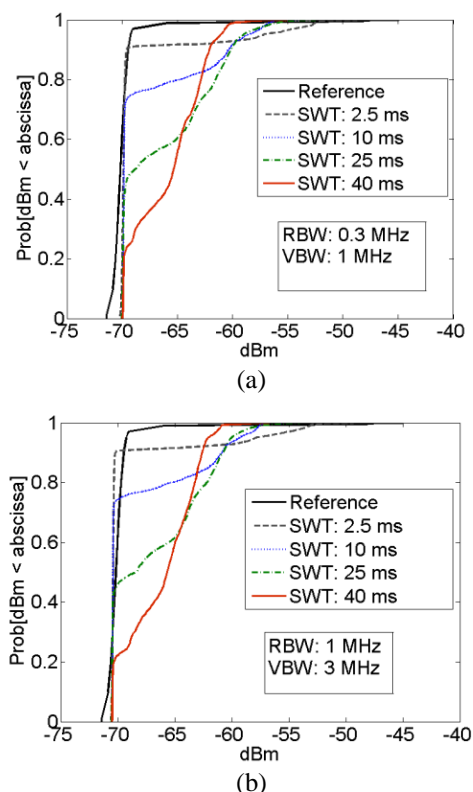

(a)

(b)

Fig. 5. CDFs of the power levels obtained from the measurements carried out in the time and frequency domains.

From the shape of the previous curves, it was determined that data collected when using higher SWT values (25ms, 40ms) would account for an idle signal with lower peaks but longer-lasting ones by far. This would imply an overestimation of the exposure levels to WiFi signals, and thus, shorter SWTs should be considered when the purpose is to maximize the accuracy of the measurements. Specifically, the tests performed with a SWT of 2.5 ms fitted better the trend of the reference curve, and thus, the use of this value leads to more rigorous and realistic results. The error of the samples taken in the frequency domain with this specific sweep time value was calculated by using equation (2), considering the 50th percentile of the measurement levels. Values that ranged between 3.40% and 9.06% and an average error equal to 5.73% were obtained when selecting a RBW of 0.3 MHz. However, the use of the same RBW value with a SWT equal to 10 ms led to values of the error between 0.20% and 12.94%, being the average error 7.80% in this case. This error increased for a SWT of 40 ms reaching a value of 203.94%.

Apart from the previous curves and errors, statistical results of second order were also determined taking into account fifteen measurements performed with each spectrum analyzer configuration. The mean, maximum and minimum values calculated from the 50th percentiles (P50) of the power levels obtained in the frequency domain were compared with the P50 of the reference values measured over time. As observed in Fig. 6, the P50 value calculated from the measurements carried out in the frequency domain when selecting a resolution bandwidth of 1 MHz, a video bandwidth of 3 MHz and sweep times of 2.5 ms or 10 ms, is lower than the median of the values recorded in the time domain ("Reference"). Thus, the WiFi exposure would be underestimated if those settings are used. For a RBW of 0.3 MHz and a VBW of 1 MHz, the statistical values corresponding to the samples taken in the frequency domain were slightly higher than the ones calculated from the time domain recordings, in 14 out of the 15 measurements carried out with a SWT of 10 ms, and for all the measurements performed with a SWT of 2.5 ms. As mentioned before, the highest error in these two specific cases was respectively 12.94% and 9.06%, concluding that these were the most suitable configurations for assessing the WiFi exposure caused by idle mode signals.

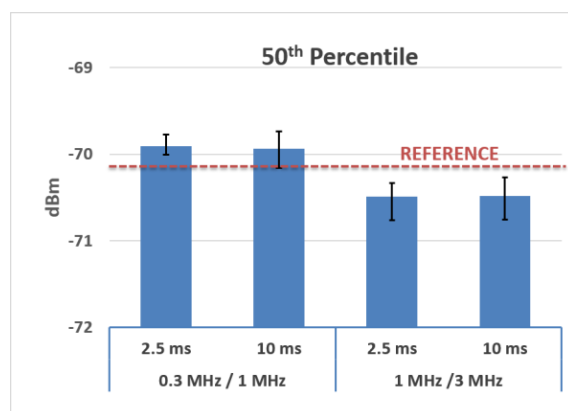

Fig.6. P50 values of both, the levels measured using different configurations in the frequency domain and the reference values obtained in the time domain.

B. *Results obtained for different WiFi data traffic situations*

Results obtained from recordings of 6-minute duration performed in the frequency domain while generating data traffic are given below. In this case, the measurements were carried out using the two configurations of the spectrum analyzer that were found to be more suitable for assessing the WiFi exposure caused by idle mode signals, because a setup that is capable of determining real exposure values derived from very low WiFi activity levels is also adequate for measuring WiFi signal bursts that imply higher percentages of reception. Thus, CDFs of the received signal levels when using a resolution bandwidth of 0.3 MHz, a video bandwidth of 1 MHz and sweep times of 2.5 ms and 10 ms are shown in Fig. 7.

values of 20 measurements registered for each type of file and measurement configuration. This means a total of 120 frequency domain measurements for obtaining the results of Table II and Table III. In these tables, the mean of the maximum power values have been also included for each case.

TABLE II
POWER LEVELS MEASURED FOR DIFFERENT
WIFI DATA TRAFFIC SITUATIONS USING A SWT OF 2.5 MS

| File | Relevant Percentiles | Mean (dBm) | Range of values (dBm) |
|---|---|---|---|
| File 1 | P90 | -65.6 | -69.3 / -59.7 |
|  | P97 | -44.7 | -47.7 / -43.3 |
|  | Max | -39.9 | -43.4 / -37.2 |
| File 2 | P80 | -60.3 | -69.2 / -51.7 |
|  | P85 | -44.6 | -46.4 / -43.3 |
|  | Max | -37.2 | -38.8 / -36.7 |
| File 3 | P30 | -66.8 | -69.1 / -53.5 |
|  | P40 | -47.7 | -50.4 / -42.2 |
|  | Max | -38.1 | -38.4 / -36.9 |

TABLE III
POWER LEVELS MEASURED FOR DIFFERENT
WIFI DATA TRAFFIC SITUATIONS USING A SWT OF 10 MS

| File | Relevant Percentiles | Mean (dBm) | Range of values (dBm) |
|---|---|---|---|
| File 1 | P70 | -67.4 | -69.2 / -61.6 |
|  | P90 | -50.9 | -52.7 / -49.9 |
|  | Max | -40.2 | -44.4 / -37.9 |
| File 2 | P60 | -68.7 | -69.5 / -65.9 |
|  | P80 | -50.0 | -52.6 / -48.9 |
|  | Max | -37.5 | -38.1 / -37.2 |
| File 3 | P30 | -64.4 | -66.1 / -63.4 |
|  | P40 | -47.4 | -49.3 / -45.5 |
|  | Max | -38.7 | -38.8 / -38.5 |

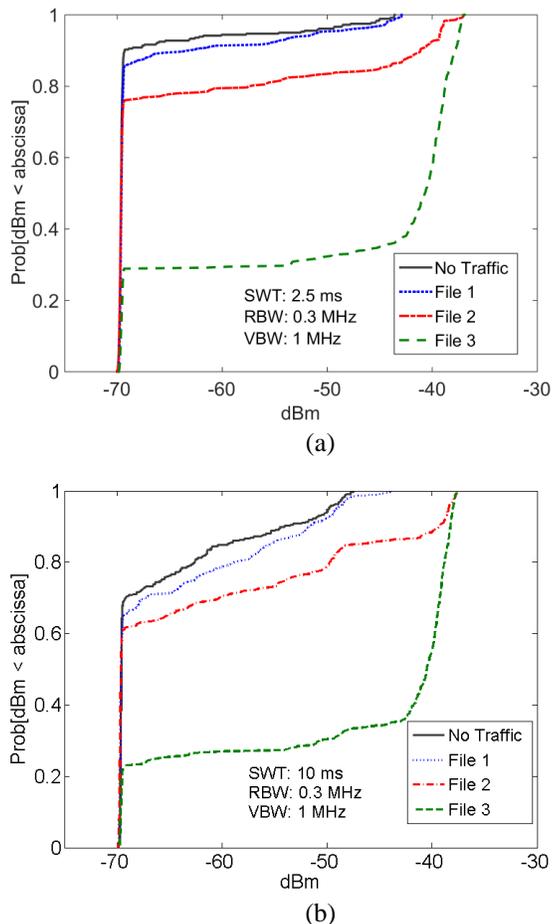

Fig. 7. CDFs of the power levels measured for different data traffic situations using RBW=0.3 MHz, VBW=1 MHz and (a) SWT=2.5 ms, (b) SWT=10 ms.

According to the percentages of WiFi activity indicated at the end of Section III, the minimum power levels associated with the transmission/reception of WiFi data traffic should be those corresponding to the $97^{th} - 98^{th}$ percentiles in the case of File 1, the $82^{th} - 85^{th}$ percentiles for File 2 and the $32^{th} - 36^{th}$ percentiles for File 3. This implies that rough transitions from the lowest to the highest power values of the corresponding CDF curves should happen at those percentiles. Thus, comparing the results of Fig. 7a and Fig. 7b, it is concluded again that a SWT of 2.5 ms provides more accurate results than a SWT of 10 ms, since in this last case, the transitions happen at lower percentiles than expected, and consequently, lead to believe that the WiFi exposure is higher than it actually is.

To confirm that overestimation, different percentiles corresponding to the reception of high signal levels due to the existence of WiFi data traffic were identified, and the mean power of those percentiles was finally calculated by using the

Very similar maximum power levels were determined with both SWT values, and therefore, this parameter has little influence if the measurements are performed to characterize the worst-case traffic mode scenario. However, it turns to be relevant when WiFi signal samples are taken with the aim of doing a realistic analysis of the corresponding exposure. As seen in the previous tables, the time-periods of the power levels derived from the existence of data traffic between the transmitter and receiver match with the durations of the file downloads, when setting the SWT to 2.5 ms. Even more, the P90, P80 and P30 values corresponding to downloads of the Files 1, 2 and 3 respectively were higher for a SWT of 10 ms than for a SWT of 2.5 ms, concluding also that the lower the WiFi activity is, the worse is the overestimation due to the use of a longer sweep time.

The influence of the SWT parameter according to the amount of traffic was also assessed by applying the Analysis of Variance (ANOVA) method [36]. Table IV shows the values of the Fisher statistics (*F-value*) and the probabilities (*p-value*) of obtaining F-values lower than the critical value that were calculated from the power levels received when downloading each type of file using the selected SWTs (2.5 ms and 10 ms). To do this, the 6-minute measurements of each type were selected (for both SWT and the three types of files), and a significance level of 0.05 (α parameter) and 359 degrees of freedom (dof parameter) were considered.

TABLE IV
RESULTS OBTAINED FROM THE APPLICATION OF THE ANOVA METHOD TO DIFFERENT WIFI DATA TRAFFIC SITUATIONS

| Traffic Situation | Percentage of WiFi Reception | F-value | p-value |
|---|---|---|---|
| Download of File 1 | 2-3 | 2.1863 | 0.0000 |
| Download of File 2 | 15-18 | 1.3904 | 0.0009 |
| Download of File 3 | 64-68 | 1.1643 | 0.0750 |

According to the results of Table IV, the lower the probability is, the more influence of the SWT on the measurements, and thus, it is concluded again that the overestimations due to the use of longer sweet times will be more critical if the WiFi activity levels are very low, as in the case of File 1. Therefore, once demonstrated that the use of a short SWT increases the accuracy of the results, it can be stated that the optimal spectrum analyzer configuration for obtaining realistic values of the WiFi exposure only and exclusively in the frequency domain is the one indicated in the following table.

TABLE V
OPTIMAL SPECTRUM ANALYZER CONFIGURATION FOR MEASURING REALISTIC WIFI EXPOSURE VALUES

| Parameter | Value |
|---|---|
| $f_c$ | Central frequency of the channel |
| Span | 20 MHz |
| RBW | 0.3 MHz |
| VBW | 1 MHz |
| SWT | 2.5 ms |
| SWP | 501 points |
| Detector | RMS |
| Trace Mode | Clear/Write |

As described in the IEEE Standard 802.11 [31], all the signals transmitted in the 2.4 GHz WiFi band use the same time and frequency masks, and thus, the settings specified in Table V can be applied directly to perform WiFi exposure measurements in any of the 20-MHz bandwidth channels defined in that band. By means of the frequency domain measurements, an exposure value per second can be obtained when performing just one measurement and this is the main advantage over the time domain recordings, which require several measurements.

V. PROCEDURE FOR ASSESSING WIFI EXPOSURE

As stated in the previous sections, it is essential to utilize the spectrum analyzer optimal settings for acquiring WiFi radiation samples, when the purpose of the measurements is to know the actual exposure levels caused by this type of signals in a particular environment. Thus, the work and results so far described in this paper were focused on identifying such configuration.

Nevertheless, another key question that should be taken into account is the measurement procedure in which that configuration will be adopted. Therefore, a set of guidelines to carry out WiFi exposure recordings is given below, based not only on the use of the spectrum analyzer setup indicated in Table V, but also on the general recommendations of several standards defined in this regard [6], [7], [8], measurement campaigns performed by other authors and on the authors' experience.

First, the antenna that will be connected to the spectrum analyzer must be chosen. When the aim of the measurements is to determine the human exposure levels caused by WiFi radiation, an isotropic or a tri-axial antenna should be used. If this is not possible, three exposure samples can be respectively taken in the x, y and z spatial directions, in order to register separately three mutually orthogonal components of the electric field received at a specific point [8].

The following step is to select the measurement locations within the area of study, taking into account the potential spatial variability of the WiFi signals, and ensuring that the receiving antenna is not in the near field region of any WiFi source. In case of performing indoor measurements, a good option is to take samples in the middle and the corners of the corresponding rooms, as well as on all those locations where people spend most of their time [17], [18]. The height of the receiving antenna should be defined in accordance with the height and the position of the individuals who are usually present in such environment (e.g. head location when they are sitting or standing), bearing also in mind that a maximum height of 2 m above the floor is recommended by the Institute of Electrical and Electronics Engineers [7]. Besides, the recordings should be done at enough distance from the walls in order to avoid their influence, e.g. a distance equal to 1.4 m, as reported in [17].

Apart from the previous recommendations, the use of a software tool developed to control the equipment and save the results is suggested in order to avoid the influence of the person who performs the measurements. This tool should be programmed to take samples of at least 6-minute duration [6], or even of 24-hours duration in the specific case of acquiring data to analyze the time variability of the exposure to WiFi signals. An example of this type of variability is shown in Fig. 8. The curves depicted in this figure account for the maximum values and the 50[th] and 90[th] percentiles of the WiFi signal levels registered inside a classroom a of the University of the Basque Country during a whole working day. To do this, an automated measurement system composed of a tri-axial antenna and a spectrum analyzer configured according to the settings indicated in Table V were used. The central frequency selected to perform the measurements was the one corresponding to the operation channel of the access point located in that classroom (Channel 5 of the 2.4 GHz band). Also, the antenna was placed in the middle of the area under study at a height of 1.2 m above the floor, since people usually sit in that specific location. As expected, the radiation was significantly higher when the university, and therefore the classroom, were open. That is, from 7:00 to 21:00.

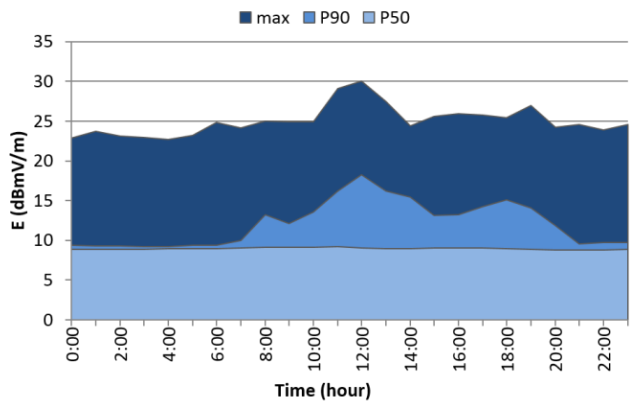

Fig. 8. Time variability of the WiFi radiation measured in a classroom of the University of the Basque Country.

## VI. CONCLUSIONS

Although human exposure to WiFi signals is nowadays a matter of social concern, there is no standardized procedure for measuring this type of emissions. Moreover, many of the scientific studies performed in this regard do not take into account that the WiFi radiation is not received in a continuous way. In fact, the most typical instruments utilized to acquire exposure samples, such as broadband probes, exposimeters or spectrum analyzers are usually configured, or even developed, to measure the maximum field levels. This can lead to significant overestimations that could finally result in a characterization of the exposure not matching the real environment conditions. For this reason, some authors weight the samples by applying empirical factors that account for the quasi-stochastic nature of the WiFi signal. However, in those cases, measurements should be done in both, the time and frequency domain.

Taking into account this, and after concluding that spectrum analyzers using a RMS detector and the Clear/Write trace mode are the most suitable instruments to measure the actual WiFi exposure levels, authors of this paper defined a methodology to identify the optimal measurement setup to acquire accurate samples of the WiFi radiation only and exclusively in the frequency domain. In fact, the technique adopted to do this can be applied to know the most appropriate setup for registering any type of radiation, since it is based on utilizing a set of reference values from recordings of a well-known signal transmitted by any source radiating the type of emissions under study.

The reference data specifically used in this work were obtained from time domain measurements of the idle mode signal transmitted by a WiFi access point working in the 2.4 GHz band. From their comparison with the power levels recorded for the same type of signal by configuring the spectrum analyzer in the frequency domain to operate at the central frequency of the transmission/reception channel, with a Span of 20 MHz, the RMS detector and Clear /Write Trace mode and a SWP value of 501, it was concluded that the optimal SWT, RBW and VBW were 2.5 ms, 0.3 MHz and 1 MHz respectively. Other configurations led to higher overestimations, and even in some cases, to underestimations that gave as a result unusable power values to determine if the WiFi radiations sources fulfill the protection thresholds or not.

The signal levels registered during a second set of frequency domain tests carried out for different traffic situations proved also that the above mentioned configuration can be used to measure accurately the exposure derived from low, medium and high WiFi activity levels. Even more, the time variability of the WiFi radiation that was received during a whole working day in a classroom of the University of the Basque Country was determined from recordings performed by applying a measurement methodology in which that configuration was adopted. Therefore, having concluded that the solution proposed by the authors is optimal for obtaining realistic values of the WiFi exposure at different environments and under different circumstances, further work will be aimed to employ not only the optimal settings here described, but also the methodology defined to determine those optimal settings, in order to identify the best measurement solution to acquire samples of the human exposure levels caused by WiFi signals transmitted in the 5 GHz band, or even by other types of signals, as for example the ones corresponding to 5G mobile communication systems.

## ACKNOWLEDGEMENT

This work has been financially supported by the Basque Government (IT-683-13), and by the Spanish Ministry of Economy and Competitiveness under the project 5G-NewBROs (TEC2015-66153-P).

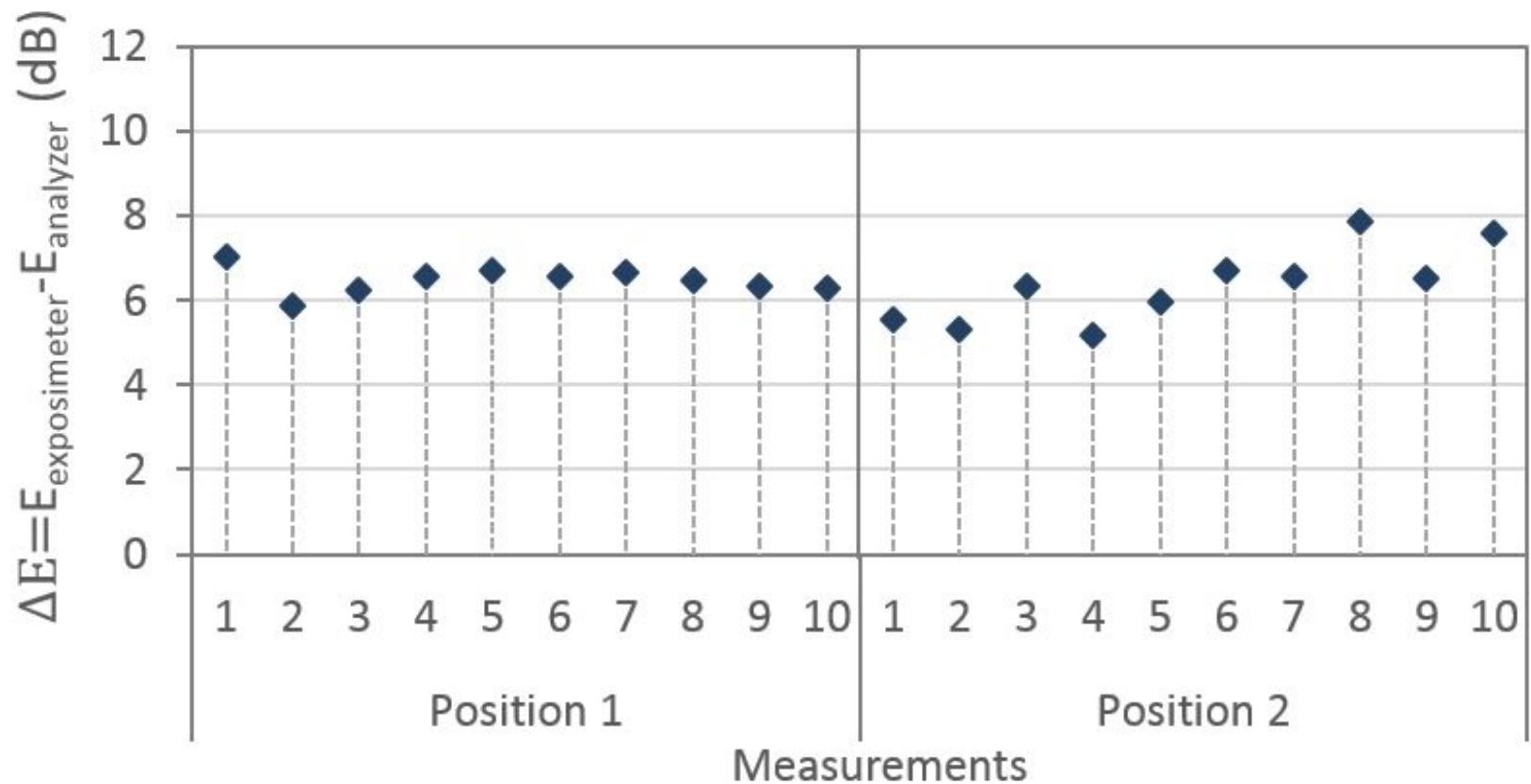

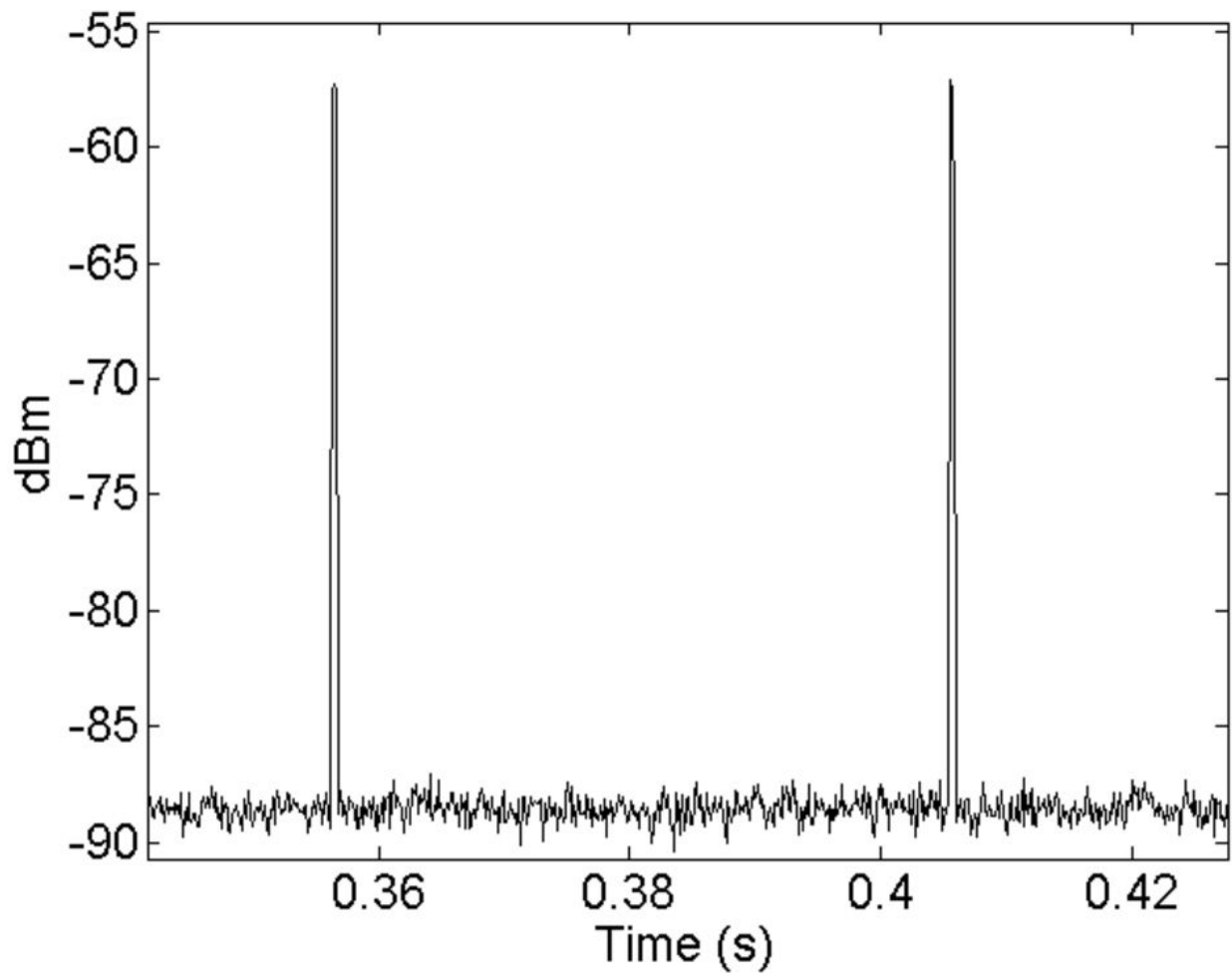

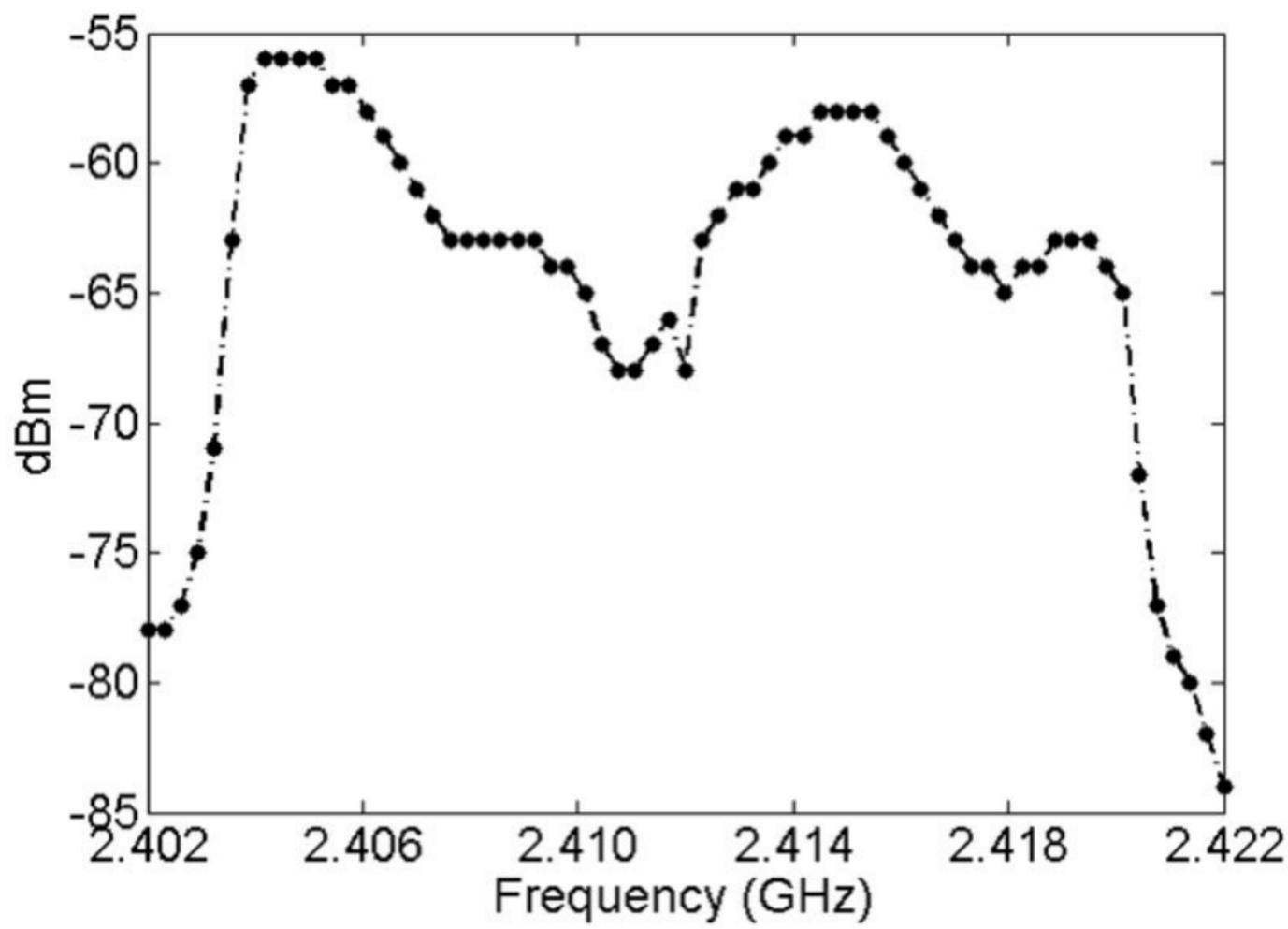

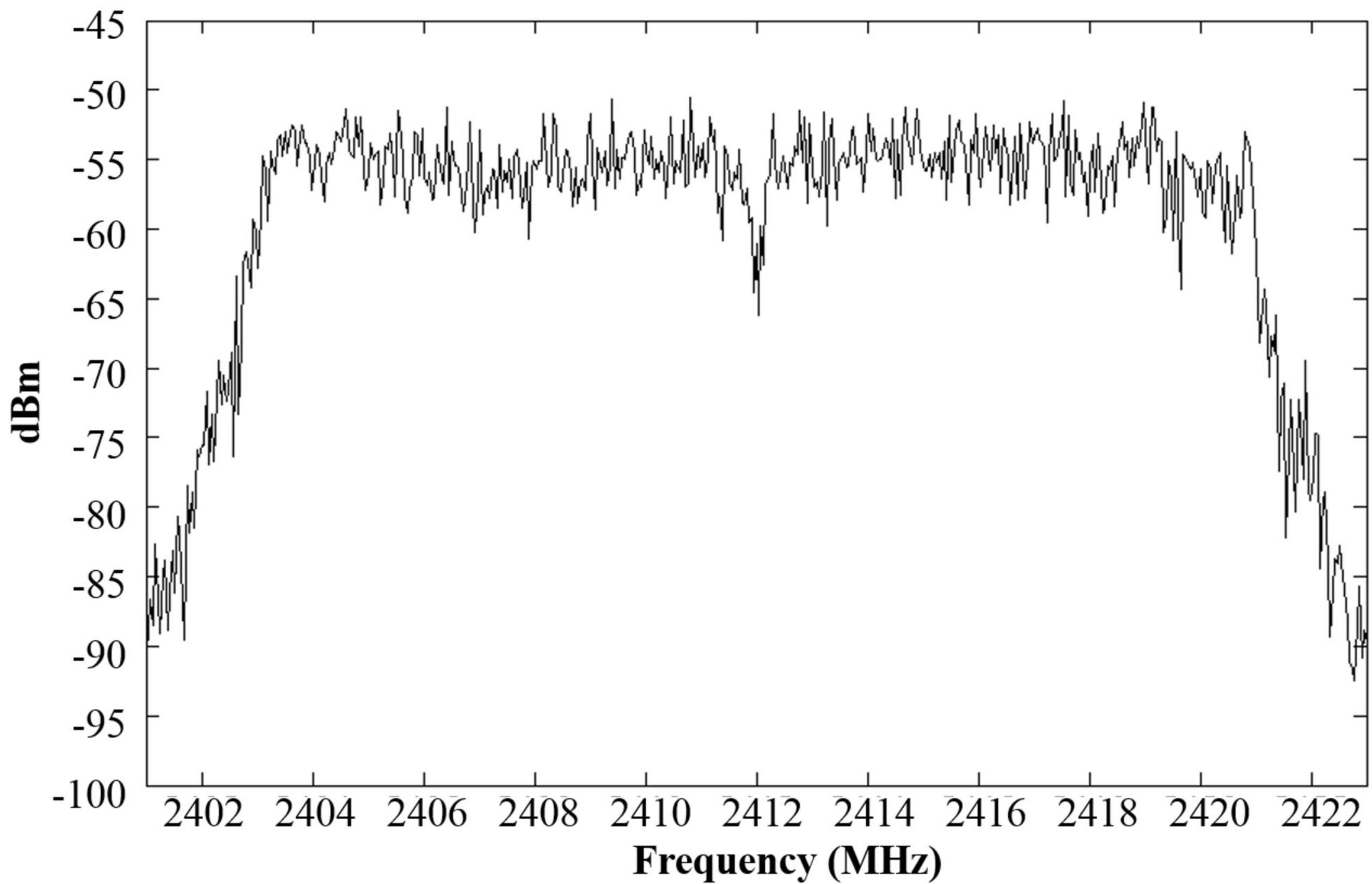

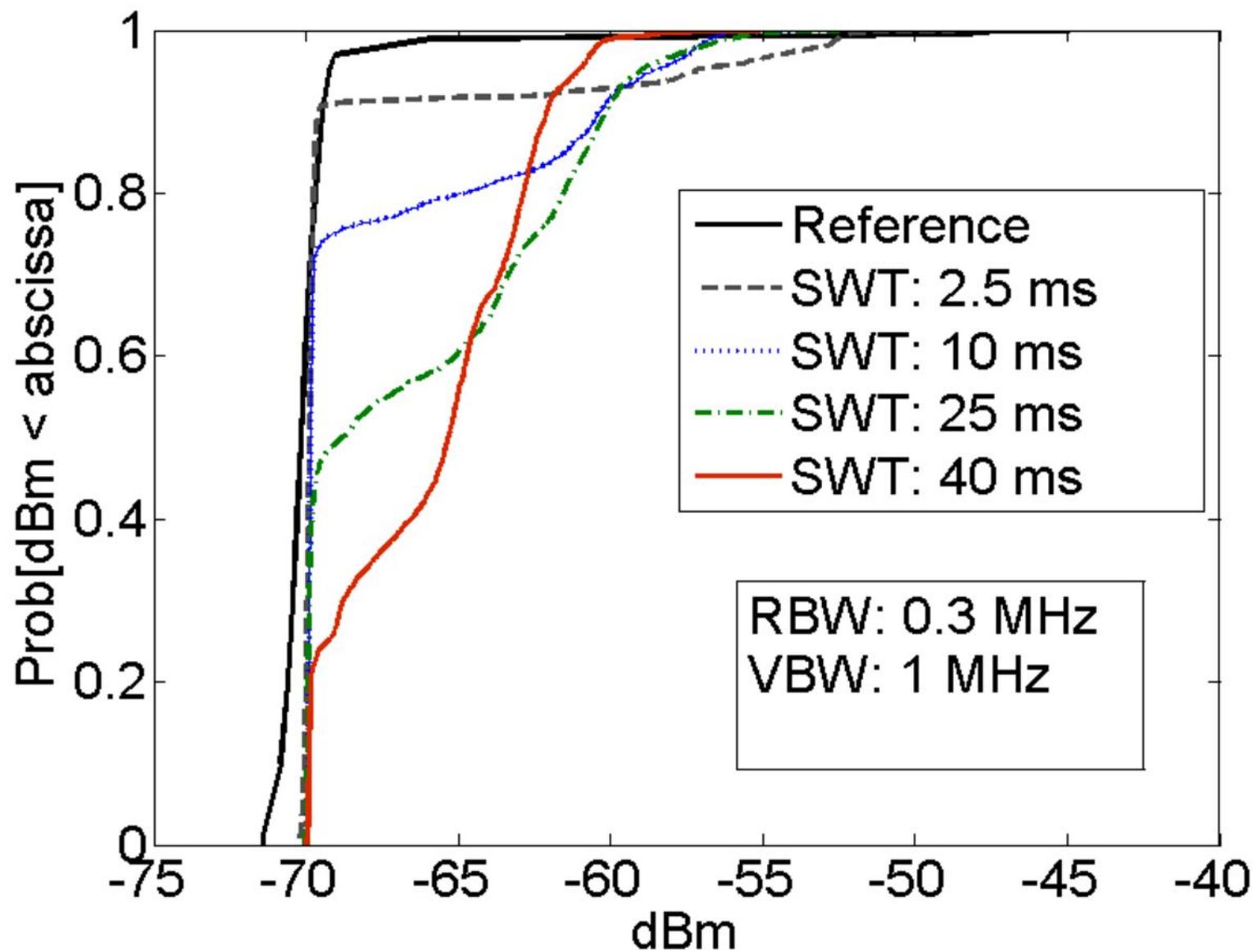

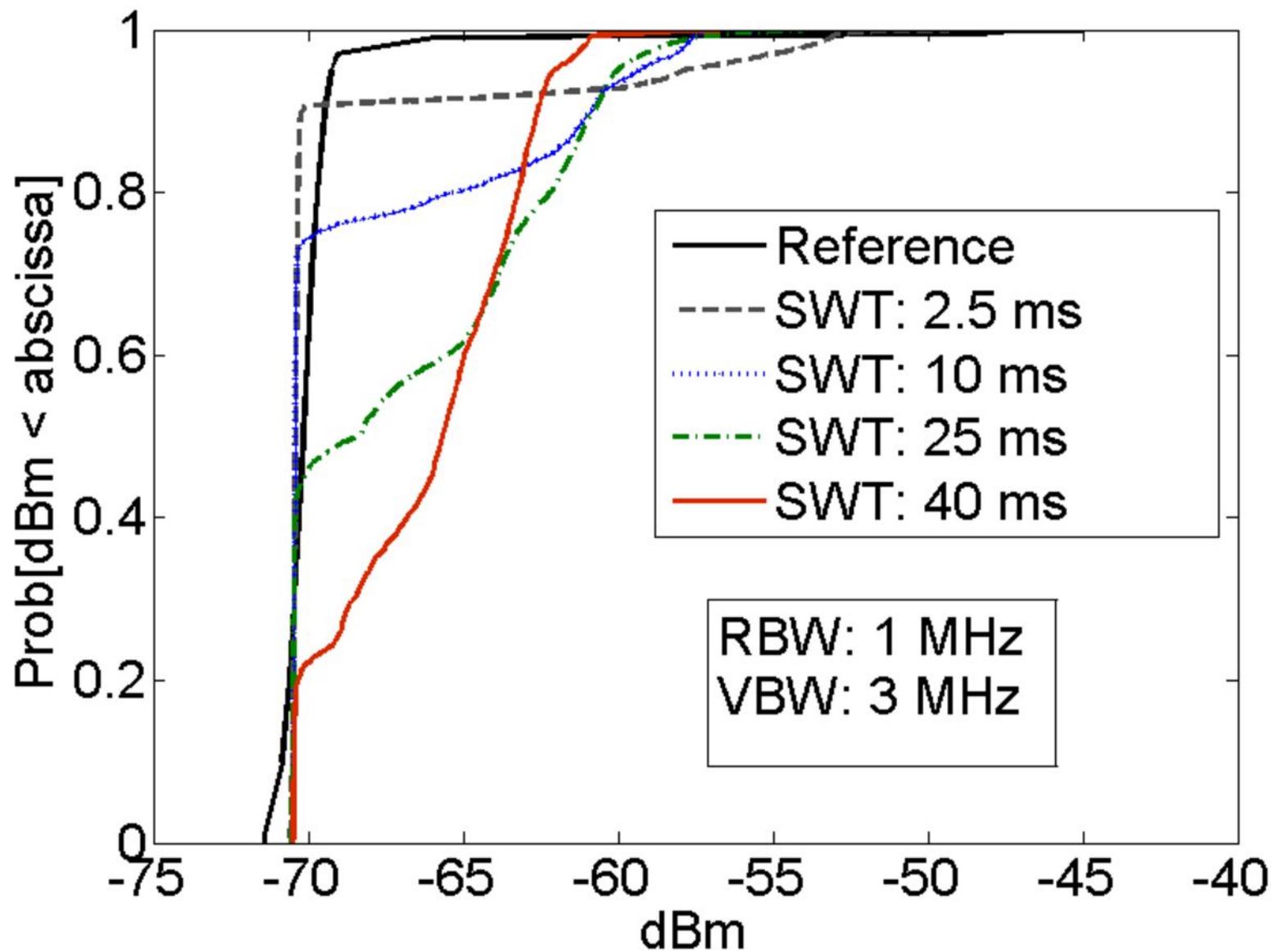

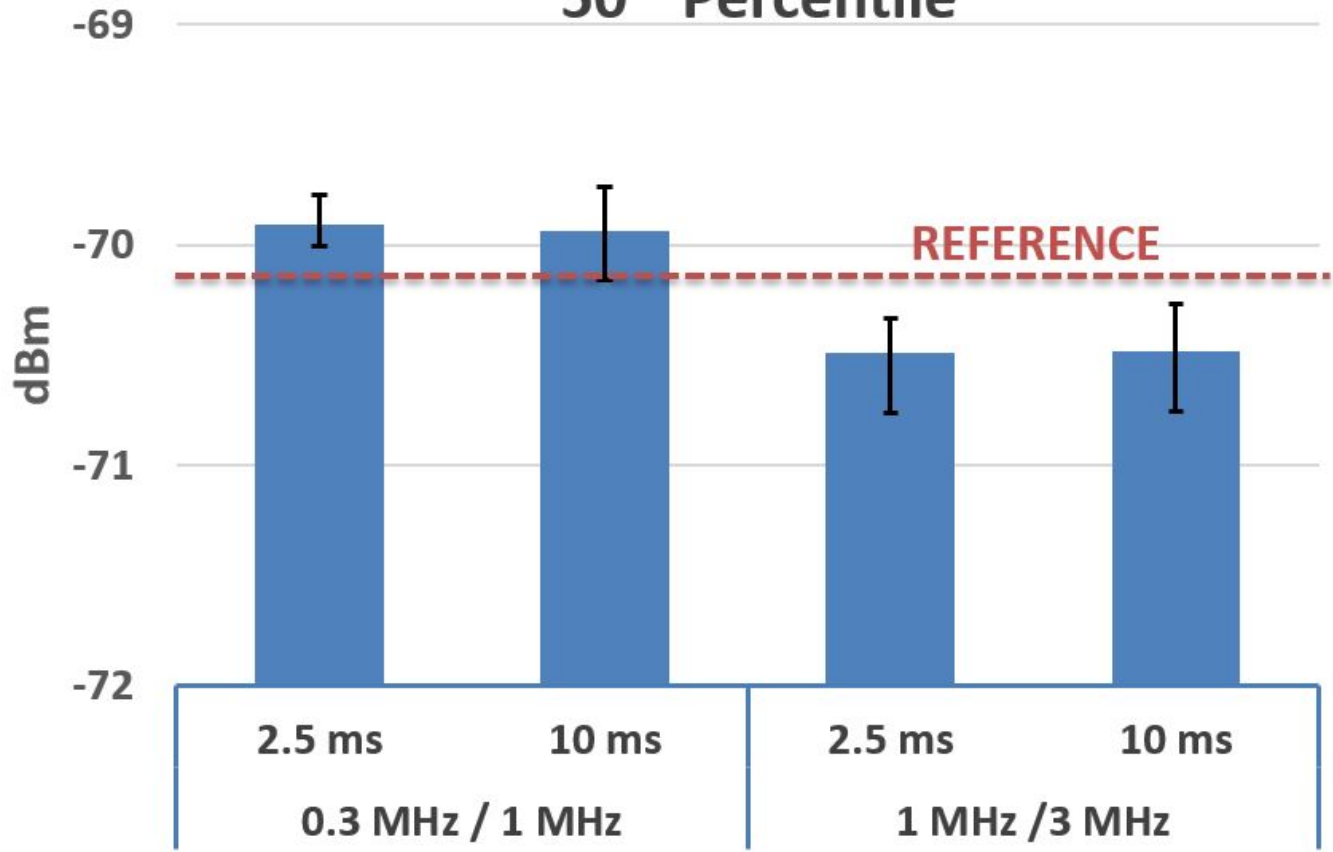

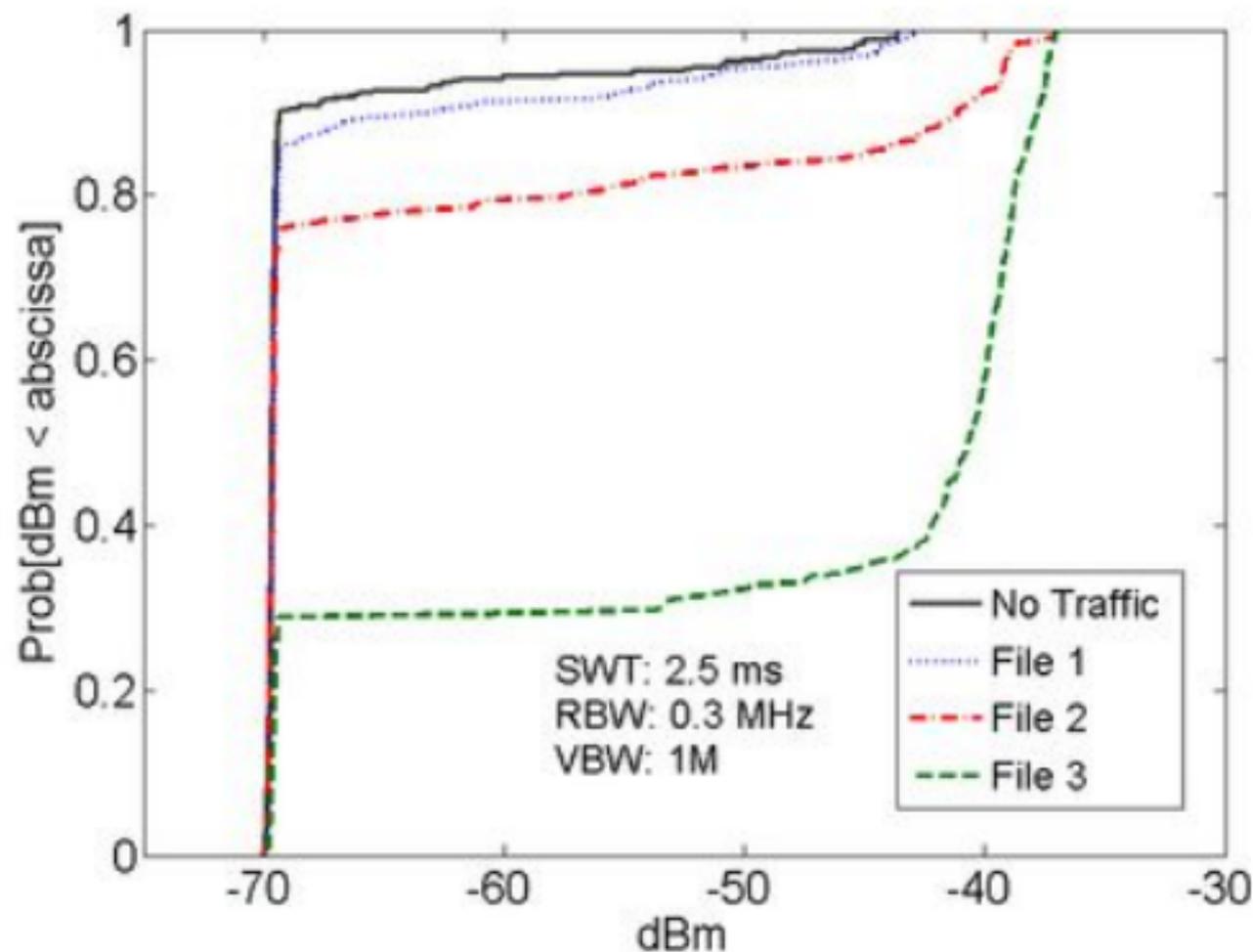

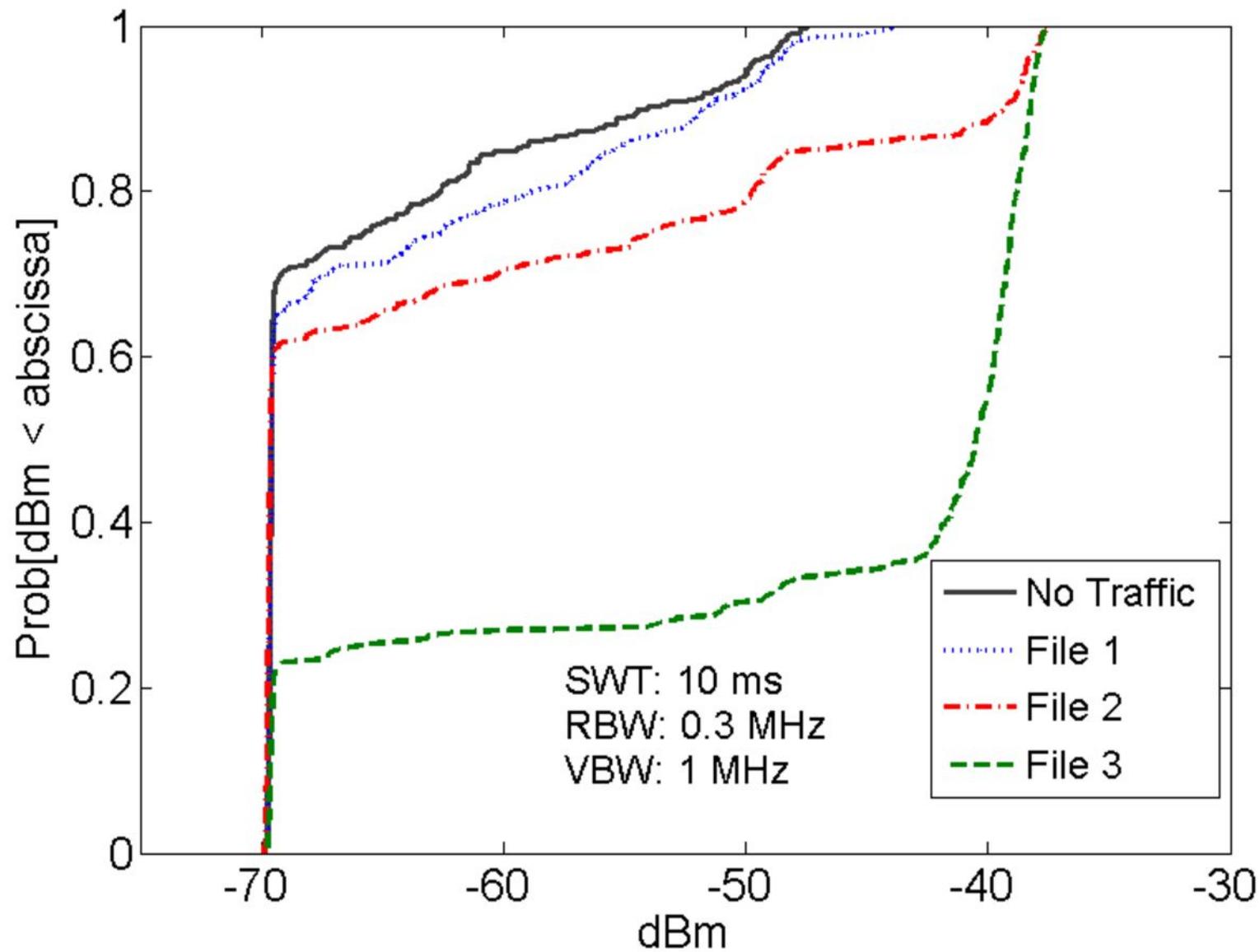

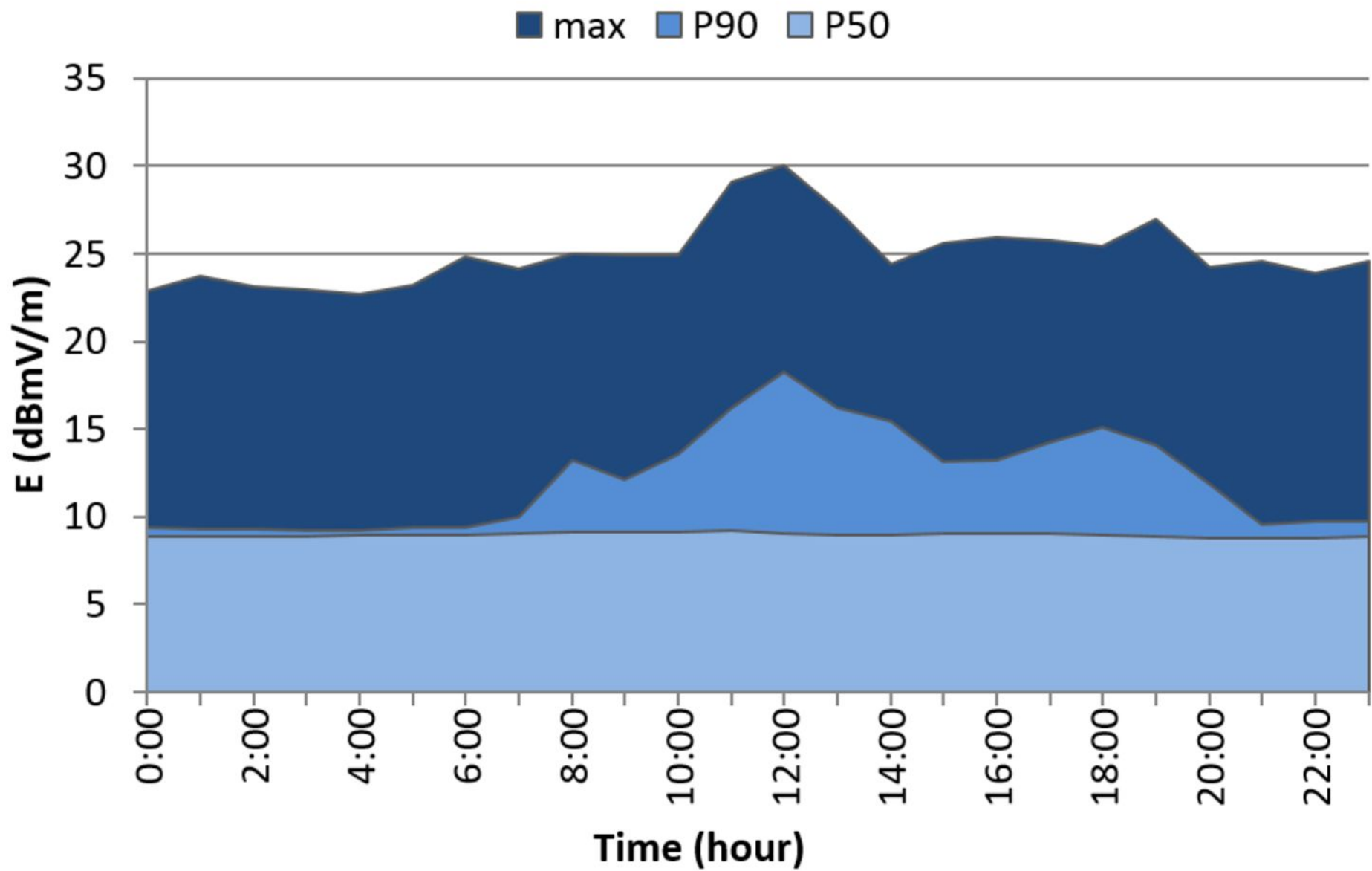

TABLE I
SPECTRUM ANALYZER CONFIGURATIONS

| Parameter | Time Domain | Frequency domain |
| --- | --- | --- |
| $f_c$ (MHz) | $2412 \pm 0.3125N$<br>$N = 0, 1, 2, \ldots 32$ | 2412 |
| Span (MHz) | Zero Span | 20 |
| RBW (MHz) | 0.3 | 0.3 - 1 |
| VBW (MHz) | 1 | 1 - 3 |
| SWT (s) | 1 | $2.5 \times 10^{-3}$ - $40 \times 10^{-3}$ |
| SWP | 8001 points | 501 points |
| Detector | RMS | RMS |
| Trace Mode | Clear/Write | Clear/Write |

TABLE II
POWER LEVELS MEASURED FOR DIFFERENT
WIFI DATA TRAFFIC SITUATIONS USING A SWT OF 2.5 MS

| File | Relevant Percentiles | Mean (dBm) | Range of values (dBm) |
| --- | --- | --- | --- |
| File 1 | P90 | -65.6 | -69.3 / -59.7 |
| | P97 | -44.7 | -47.7 / -43.3 |
| | Max | -39.9 | -43.4 / -37.2 |
| File 2 | P80 | -60.3 | -69.2 / -51.7 |
| | P85 | -44.6 | -46.4 / -43.3 |
| | Max | -37.2 | -38.8 / -36.7 |
| File 3 | P30 | -66.8 | -69.1 / -53.5 |
| | P40 | -47.7 | -50.4 / -42.2 |
| | Max | -38.1 | -38.4 / -36.9 |

TABLE III
POWER LEVELS MEASURED FOR DIFFERENT
WIFI DATA TRAFFIC SITUATIONS USING A SWT OF 10 MS

| File | Relevant Percentiles | Mean (dBm) | Range of values (dBm) |
| --- | --- | --- | --- |
| File 1 | P70 | -67.4 | -69.2 / -61.6 |
| | P90 | -50.9 | -52.7 / -49.9 |
| | Max | -40.2 | -44.4 / -37.9 |
| File 2 | P60 | -68.7 | -69.5 / -65.9 |
| | P80 | -50.0 | -52.6 / -48.9 |
| | Max | -37.5 | -38.1 / -37.2 |
| File 3 | P30 | -64.4 | -66.1 / -63.4 |
| | P40 | -47.4 | -49.3 / -45.5 |
| | Max | -38.7 | -38.8 / -38.5 |

TABLE IV
RESULTS OBTAINED FROM THE APPLICATION OF THE ANOVA METHOD
TO DIFFERENT WIFI DATA TRAFFIC SITUATIONS

| Traffic Situation | Percentage of WiFi Reception | F-value | p-value |
| --- | --- | --- | --- |
| Download of File 1 | 2-3 | 2.1863 | 0.0000 |
| Download of File 2 | 15-18 | 1.3904 | 0.0009 |
| Download of File 3 | 64-68 | 1.1643 | 0.0750 |

TABLE V
OPTIMAL SPECTRUM ANALYZER CONFIGURATION
FOR MEASURING REALISTIC WIFI EXPOSURE VALUES

| Parameter | Value |
|---|---|
| $f_c$ | Central frequency of the channel |
| Span | 20 MHz |
| RBW | 0.3 MHz |
| VBW | 1 MHz |
| SWT | 2.5 ms |
| SWP | 501 points |
| Detector | RMS |
| Trace Mode | Clear/Write |